\newcommand{\un}{~\mathrm}
\newcommand{\ie}{{\em i.e. }}
\newcommand{\eg}{{\em e.g. }}
\newcommand{\unm}{~\mu\mathrm{m}}
\newcommand{\indice}[1]{\textnormal{\scriptsize{#1}}}
\newcommand{\indicito}[1]{\textnormal{\tiny{#1}}}
\documentclass[twocolumn
,prl
,floatfix
,showpacs
,preprintnumbers
,amsmath
,amssymb
]{revtex4}

\usepackage{graphicx}
\usepackage{dcolumn}
\usepackage{bm}

\begin{document}
\title{Depinning transition in failure of inhomogeneous brittle materials}
\author{Laurent Ponson}\email{ponson@caltech.edu}
\affiliation{Laboratorio d'Estruturas e Materiais, COPPE/Universidade federal do Rio de Janeiro, CEP 21945-970 RJ, Rio de Janeiro, Brazil. \\ Division of Engineering and Applied Science, California Institute of Technology, Pasadena, CA 91125, USA.}

\begin{abstract}
The dynamics of a crack propagating in an elastic inhomogeneous material is investigated. The variations of the average crack velocity with the external loading are measured for a brittle rock and are shown to display two distinct regimes: Below a given threshold $G_\indice{c}$, the crack velocity is well described by an exponential law $v \simeq e^{-\frac{C}{G- \langle \Gamma \rangle }}$ characteristic of subcritical propagation, while for larger values of the driving force $G > G_\indice{c}$, the velocity evolves as a power law $v \simeq (G-G_\indice{c})^\theta$ with $\theta = 0.80 \pm 0.15$. These results can be explained extending the continuum theory of Fracture Mechanics to disordered systems. In this description, the motion of a crack is analogue to the one of an elastic line driven in a random medium and critical failure occurs when the loading is sufficiently large to depinne the crack front from the heterogeneities of the material.
\end{abstract}

\pacs{62.20.Mk, 
46.50.+a, 
68.35.Ct 
}
\date{\today}
\maketitle

Failure of inhomogeneous materials has been a very active field of research during the last decades (see Ref. \cite{Alava2} for a recent review). A great research effort in this field has been dedicated to the study of fluctuations: Fluctuations of velocity around the average motion of cracks when the studies were devoted to their highly intermittent dynamics \cite{Maloy2, Maloy3, Koivisto, Davidsen}, or variations to a straight trajectory when the works were dedicated to the rough geometry of fracture surfaces \cite{Bouchaud9, Maloy, Ponson5}. In both cases, these fluctuations were shown to display remarkably robust properties suggesting that crack propagation in disordered systems could be described on a general manner by relatively simple statistical models able to capture the competition between the two antagonist effects occurring during failure of inhomogeneous materials: Disorder and elasticity.

Very recently, main statistical features of fluctuations of both trajectory and velocity for cracks propagating in brittle materials were captured by stochastic models of elastic lines driven in random media \cite{Bonamy2, Bonamy5} that mimic the motion of cracks through the microstructural disorder of materials. However, the relevance of this theoretical framework for fracture problems is still a matter of debate: On the one hand, the ability of these models to describe the {\em average behavior} of the crack such as its mean velocity, or the critical external loading at failure, more interesting from a mechanical or an engineering point of view, is still an open question. On the other hand, a direct experimental observation of the {\em critical dynamic transition} from a crack pinned by the heterogeneities of the material ($v=0$) to a propagating crack ($v>0$), as predicted by this theory at the onset of material failure (driving force $G=G_\indice{c}$), is still lacking. The investigation of this depinning transition on an experimental example is the central point of this Letter.

The variations of the average crack velocity with the external driving force, \ie the energy release rate $G$ \cite{Lawn}, are measured for a brittle rock. They are shown to exhibit two distinct regimes. Below a critical threshold $G_\indice{c}$, the crack velocity is well described by an exponential law $v \simeq e^{-\frac{C}{G- \langle \Gamma \rangle }}$ characteristic of a subcritical propagation, while for larger values of the external loading $G > G_\indice{c}$, the velocity evolves as a power law $v \simeq (G-G_\indice{c})^\theta$ with $\theta = 0.80 \pm 0.15$. This behavior is fully captured by a stochastic model directly derived from Fracture Mechanics extended to inhomogeneous systems where crack propagation is analogue to the motion of an elastic line driven in a random medium \cite{Schmittbuhl4, Ramanathan, Bonamy5}.

{\em System and setup.} -

Sandstone is chosen as an archetype of heterogeneous elastic materials. A Botucatu sandstone, extracted in the central region of Brazil, has been used for the experiments. It is made of quartz grains with a diameter $d = 230\unm \pm 30\unm$ and a porosity $\phi = 18 \pm 2\un{\%}$, that results in highly inhomogeneous mechanical properties at the grain scale. This South-American rock is consolidated thanks to an iron oxide cement providing to the rock a red coloration. As a result, its fracture energy $G_\indice{c} \simeq 140\un{J.m^{-2}}$ as measured in the following is relatively high compared to other sandstones \cite{Atkinson}. Its intrinsic tensile strength measured by splitting cylinders submitted to uniaxial compression \cite{Carneiro2} is found to be $\sigma_\indice{Y} = 75\un{MPa} \pm 5\un{MPa}$, leading to an estimate of its process zone size $\ell_\indice{PZ} \simeq \frac{\pi}{8} \frac{G_\indicito{c} E}{\sigma_\indicito{Y}^2}$ \cite{Barenblatt} \--- zone next to the crack tip where damage and dissipative processes are localized during failure \--- of $\ell_\indice{PZ} \simeq 200\unm$. This failure mechanisms property ($\ell_\indice{PZ} < d$) suggests that crack propagation in the Botucatu rock mimics the crack motion in ideal brittle materials at the scale of the quartz grains.

\begin{figure}
\includegraphics[width=1.\columnwidth]{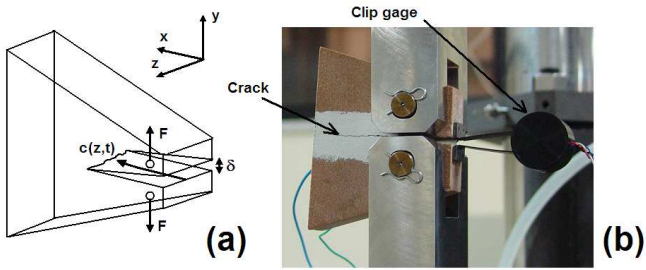}
\centering
\caption{Experimental setup. (a) Sketch of the Tapered Double Cantilever Beam  geometry; (b) picture of the specimen during crack propagation.}
\label{Fig1}
\end{figure}

A new experimental setup has been developed in order to measure variations of crack velocity from slow to very fast propagation in brittle materials. Contrary to various fracture tests as the Double Torsion geometry \cite{Evans} classically used to measure loading/crack velocity relationship in rocks, the Tapered Double Cantilever Beam specimens used in the experiments (Fig.\,\ref{Fig1}(a)) result in a slight, but controlled acceleration of the crack that is produced by the tapered shape of the specimen. As a consequence, it is possible to measure crack velocities up to $v \simeq 1\un{m\,s^{-1}}$ not achieved by classical tests used in Rock Mechanics \cite{Atkinson, Nara}. In addition, failure is obtained by the propagation of a straight crack front in the specimen, so that the local velocity at each position of the crack line is also the average velocity of the front, allowing for great simplifications in the theoretical analysis of the experiments. Finally, a straight crack propagation in the specimen is obtained without any lateral guide grooves that are known to induce systematic errors on experimental $v(G)$ curves \cite{Nara}.

\begin{figure}
\includegraphics[width=0.9\columnwidth]{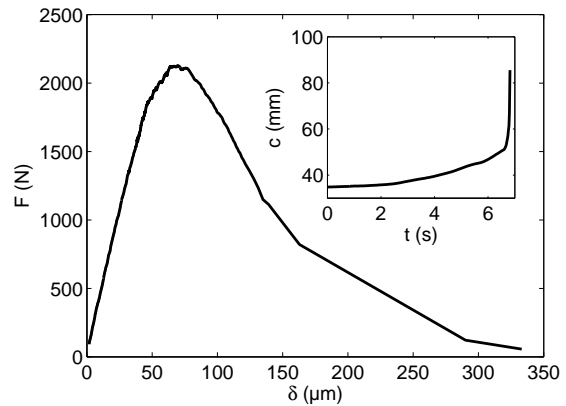}
\centering
\caption{Mechanical behavior of the specimen. (a) Typical load-crack opening displacement curve; (b) corresponding evolution of the average position of the crack front.}
\label{Fig2}
\end{figure}

An initial notch $c_0 = 35\un{mm}$ is machined in $100\un{mm}$ long samples with thickness $e=30\un{mm}$. They are then submitted to a uniaxial traction by increasing the displacement $\delta_\indice{F} = v_\indice{ext} t$ between two rods inserted in the drilled specimens at constant velocity $0.2\unm\,\un{s}^{-1} \leq v_\indice{ext} \leq 4\unm\,\un{s}^{-1}$. During the test, a gauge force measures the applied tension $F$ while a clip gage measures the opening displacement $\delta$ between the two lips of the crack with a precision of $0.1\unm$ (see Fig.\,\ref{Fig1}(b)). A typical force-crack opening displacement curve obtained during a fracture test of a Botucatu specimen at room temperature is presented in Fig.\,\ref{Fig2}.

The initial linear part of the curve \--- prior crack initiation \--- allows for an estimation of the Young's modulus $E = 25 \pm 1 \un{GPa}$ of the sandstone, in agreement with the value obtained from measurements of its compressive and shear waves speed. After crack initiation, the average position of the crack front $c = \langle c(z) \rangle_z$ is measured using Finite Element (FE) simulations of an elastic specimen in the same geometry. The numerical compliance $\lambda^\indice{FE}(c)$ is compared to the experimental compliance $\lambda(t) = \delta/F$ in order to get the evolution of the crack length $c(t)$ represented in inset of Fig.\,\ref{Fig2}. Other techniques limited to the free surface of the sample based either on image analysis of the crack motion at the surface or on the resistance measurement of a thin conductive film deposited on the sample side have led to similar, however less precise measurements of the crack length.

From the evolution of the crack length, it is now possible to measure the crack speed $v = \frac{dc}{dt}$ as well as the driving force $G$ imposed to the system during the test. Using the load-displacement curve to measure the work $\delta W$ of the tensile machine during the span $\delta t$\cite{Note1}, one gets $G(t) = \frac{\delta W(t)}{e [c(t+\delta t) -c(t)]}$\cite{Morel7}. On the other hand, the driving force is estimated independently using the relation $G(t) = [F(t))]^2 g^\indice{FE}[c(t)]$ where the geometrical part $g^\indice{FE}$ of the driving force is provided by the FE simulations. Both methods lead to similar results within $2\un{\%}$.

\begin{figure}
\includegraphics[width=0.9\columnwidth]{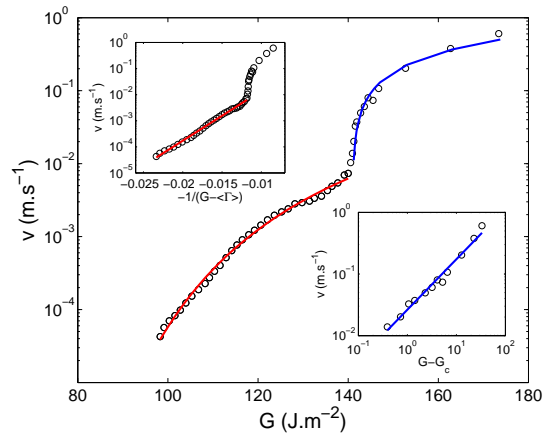}
\centering
\caption{Average dynamics of a crack propagating in Botucatu sandstone. The variations of the crack velocity are plotted in logarithmic scale with respect to the external loading. The subcritical regime $G<G_\indice{c}$  with $G_\indice{c} = 140\un{J.m^{-2}}$ is studied in the top-left inset. Solid line corresponds to the best fit of the data in $v \sim e^{-\frac{C}{G-\langle \Gamma \rangle}}$ obtained for $\langle \Gamma \rangle = 63\un{J.m^{-2}}$. Bottom right inset shows the velocity variations with the net loading $G-G_\indice{c}$ in a logarithm representation for $G>G_\indice{c}$. Straight line corresponds to a power law fit with exponent $\theta = 0.81$.}
\label{Fig3}
\end{figure}

{\em Experimental results.} -
The variations of the crack velocity with the driving force as observed on the sandstone specimens are represented in Fig.\,\ref{Fig3} in semi-logarithmic coordinates. Velocity measurements are achieved over almost five orders of magnitude, corresponding to a relatively small variation of the driving force. Irrespective of the external loading rate as well as the sample investigated, the failure behavior of the rock is found to be systematically characterized by two distinct regimes defining $G_\indice{c}$. Near, but above this critical loading, a slight change in the driving force results in a strong variation in the crack velocity. This high sensibility is studied in more detail in the bottom right inset of Fig.\,\ref{Fig3}, where $v$ is plotted as a function of the distance to the critical force $G-G_\indice{c}$ in logarithmic coordinates. The linear behavior in this representation suggests a power law variation of the crack velocity $v \sim (G-G_\indice{c})^\theta$. The value of $G_\indice{c} = 140 \pm 3 \un{J.m^{-2}}$ is found to optimize this scaling relation, and leads to an exponent $\theta = 0.80 \pm 0.15$ where the error bars are calculated from the variations measured from sample to sample.

The variations of velocity at low driving forces $G<G_\indice{c}$ are now studied. Contrary to the previous regime, slow crack propagation in rocks has been largely investigated and shown to depend crucially on the temperature \cite{Atkinson, Nara}. Analytical forms as $v \sim e^{\frac{-E^*}{k_\indicito{B} T}} G^{n/2} $ \cite{Charles} or $v \sim e^{-\frac{E_0-b G}{k_\indicito{B} T}}$ \cite{Wiederhorn2} are usually used to describe the experimental data. Both formula reproduce correctly the measurements reported here as far as $G<120\un{J.m^{-2}}$. The first one, used by many researchers because it is convenient to integrate and differentiate leads to a subcritical crack growth index $n \simeq 34$ that compares well with the other experimental findings for sandstone \cite{Atkinson}. The second formula leads to $b \simeq 0.68\,.\,10^{-20}\un{m^2}$ which is also in agreement with the other measurements made on rocks with a similar structure \cite{Nara}. This description is based on the Arrhenius law $v \sim e^{-\frac{E_\indicito{a}}{k_\indicito{B}T}}$ where the activation energy $E_\indice{a} = E_0 - b\,G$ represents the typical barrier along the energy landscape of the system tilted by the external force $G$. In this approach, the tensile force produced by the external loading on the interatomic bond just next to the crack tip favours the thermal activated processes responsible for its rupture, as \eg thermal stress fluctuations \cite{Santucci3} and chemical reactions \cite{Wiederhorn}. As a consequence, the motion of the crack front is analogue to the propagation of one point in a 2D medium. As soon as one considers a 3D inhomogeneous material, the elasticity of the crack line comes into play, and one expects a different expression of the activation energy $E_\indice{a} \sim \frac{1}{G-\langle \Gamma \rangle}$ \cite{Ponson13, Koivisto} as presented in detail in the next section. This expression provides a rather good description of the data for the full subcritical regime $G<G_\indice{c}$. The best linear regression of the function $\textnormal{ln}(v) \sim \varphi (-\frac{1}{G-\langle \Gamma \rangle})$ represented in the upper left inset of Fig.\,\ref{Fig3} is obtained for $\langle \Gamma \rangle = 63 \pm 5 \un{J.m^{-2}}$.

{\em Discussion.} - 
The observation of two very different regimes, with an exponential variation of $v$ with the external loading for $G<G_\indice{c}$ and a power law behavior for $G>G_\indice{c}$, reveals a fundamental aspect of the dynamics of crack propagating in brittle inhomogeneous medium. Let's investigate theoretically the crack motion in these systems to understand quantitatively such a behavior. As a starting point, we assume that the local velocity $v(M)$ of a point $M$ of the crack is proportional to the excess of energy $G(M)-\Gamma(M)$ locally released by the system where $\Gamma$ refers to the local fracture energy of the material. In a disordered material such as sandstone, the fracture energy can be described as a stochastic field $\Gamma(M) = \langle \Gamma \rangle + \delta \Gamma\,\eta(M)$ where $\eta$ is a short range correlated random term with zero mean value and unit second order moment. The heterogeneities of the material induce perturbations of the crack front, which is parallel in average to the $z$-axis and propagates along the $x$-axis, both in the mean fracture plane $(x,z)$ (in-plane perturbations $c(z,t)-\langle c(z,t) \rangle_z$) and in the perpendicular direction $y$ (out-of-plane perturbations $h(z,t)$). They in turn lead to variations in the local value of the external driving force $G(M)$. Interestingly, for small perturbations, $G(M)$ is only depending on the in-plane deviations of the crack front \cite{Ball, Mochvan}, and is given by $G(z,t) = G + \frac{G}{\pi} \int \frac{c(z,t)-c(z',t)}{(z-z')^2} dz'$ \cite{Rice} where $G$ refers to the macroscopic driving force applied by the tensile machine to the specimen. Using the previous expressions of the local driving force and material fracture energy, one gets the following equation of motion for a crack propagating in a brittle inhomogeneous material
\begin{equation}
\frac{\partial c}{\partial t}|_{z,t} \sim (G-\langle \Gamma \rangle) + \frac{G}{\pi} \int \frac{c(z,t)-c(z',t)}{(z-z')^2} dz' + \delta \Gamma \eta(c,h,z).
\label{Eq_motion}\end{equation}
As the out-of-plane perturbations $h$ behave independently of $c$, the stochastic term in this equation is analogue to a 2D random potential depending only on $c$ and $z$. Therefore, the crack motion is described by an equation of pinning/depinning of an elastic line driven in a random medium as previously proposed for interfacial cracks propagating in inhomogeneous weak planes \cite{Schmittbuhl4, Ramanathan}: if the external loading $G$, or driving force for unit crack length, exceeds a given threshold 
\begin{equation}
G_\indice{c} = \langle \Gamma \rangle + \pi (\delta \Gamma)^2/\langle \Gamma \rangle,
\label{Gc}\end{equation}
the crack propagates, while the front is pinned by the material heterogeneities if $G < G_\indice{c} $ \cite{Barabasi}. Above the threshold, the mean velocity $v$ of the crack front is expected to scale as $(G-G_\indice{c})^\theta$ where $\theta$ is called the velocity exponent. Ertas and Kardar have studied equation (\ref{Eq_motion}) using functional renormalization group technique \cite{Ertas2}. To first order in perturbation, they find $\theta = 0.78$. Recent direct numerical simulations resulted in $\theta \simeq 0.63$ \cite{Duemmer3}. As a consequence, the power law behavior measured experimentally with an exponent $\theta \simeq 0.80$ suggests that a {\it depinning transition} from a stable to a propagating crack as described in Eq.~(\ref{Eq_motion}) occurs at $G=G_\indice{c}$. Below the threshold at zero temperature, the external driving force is not sufficient to make the crack propagate and the crack front is pinned by the material heterogeneities. However, at finite temperature $T>0$, thermally activated processes can enable a subcritical propagation. In this regime, described by adding an annealed noise $\eta_\indice{T}(z,t)$ to Eq.~(\ref{Eq_motion}), one expects also a collective motion of the line, characteristic of glassy systems, and velocity variations are predicted to follow the so-called creep law $v \sim e^{-\left(\frac{\ell}{\pi}\right)^2 \frac{\langle \Gamma \rangle^2}{k_\indicito{B} T (G- \langle \Gamma \rangle)}}$ \cite{Feigelman, Kolton}. This expression describes rather well the experimental measurements of $v(G)$ over the whole range of external loading $G<G_\indice{c}$, leading to activation energies in the range $E_\indice{a} = \left(\frac{\ell}{\pi}\right)^2 \frac{\langle \Gamma \rangle^2}{G- \langle \Gamma \rangle}\simeq 0.20-0.30\un{eV}$, one order of magnitude larger than the thermal energy $k_\indice{B} T$. Interestingly, the expression of the activation energy provides an estimate of the topothesy \--- size of the elementary feature of the crack front for which the amplitude of its roughness compares with its length \--- $\ell=0.10\un{nm}$, compatible with the interatomic distance. Finally, let us note that using Eq.~(\ref{Gc}), the experimental values of the critical loading $G_\indice{c}$ and the average fracture energy $\langle \Gamma \rangle$ allows to estimate the normalized fluctuations $\frac{\delta \Gamma}{\langle \Gamma \rangle} =  \sqrt{\frac{G_\indice{c}-\langle \Gamma\rangle}{\pi \langle \Gamma\rangle}} = 0.62 \pm 0.06$ of fracture energy in the Botucatu rock, slightly larger but comparable with an estimate of this quantity $ \sqrt{\frac{\phi}{1-\phi}} \simeq 0.5$ corresponding to an ideal porous material made of an homogeneous solid with constant fracture energy and voids, with volumic fractions $1-\phi=0.82$ and $\phi = 0.18$, respectively.

{\em Conclusion.} - 
The average dynamics of a crack propagating in a brittle inhomogeneous rock has been investigated. The velocity variations with the external loading display two distinct regimes: above a threshold $G_\indice{c}$, $v$ evolves as a power law of the net loading $G-G_\indice{c}$ with exponent $\theta = 0.80 \pm 0.15$ while for $G<G_\indice{c}$, these variations are described by an Arrhenius law $v \sim e^{-\frac{E_\indicito{a}}{k_\indicito{B} T}}$ with an activation energy $E_\indice{a} \sim \frac{1}{G-\langle \Gamma \rangle}$. Both behaviors can be explained extending the continuum theory of fracture mechanics to inhomogeneous systems. In this description, the external loading, the crack front elasticity as well as the effect of material heterogeneities interplay leading to a macroscopic motion of the line above a critical force $G_\indice{c}$ larger than the average value of the fracture energy in the material $\langle \Gamma \rangle$. Below this critical transition, the line can propagate only at finite temperature through thermal activated processes. Therefore, the measurement at the macroscopic scale of the crack velocity variations with the external loading represents a signature of the depinning of the crack front at the scale of the material microstructure. The experimental measurements reported here represent a strong argument in favour of the pinning/depinning approach to brittle failure and open new perspectives for the prediction of macroscopic quantities such as average crack velocity and fracture energy of direct interest for Engineering and Applied Science.

The author would like to thank Dr. Guilherme Cordeiro and Ashutosh Bindal for their help in the experiments and M. Alava, K. Bhattacharya, D. Bonamy, E. Bouchaud, J.-B. Leblond, S. Morel, A. Rosso and R. Toledo for helpful discussions. Financial support from the French Ministry of Foreign Affairs through the Lavoisier Program is acknowledged.


\begin{thebibliography}{34}
\expandafter\ifx\csname natexlab\endcsname\relax\def\natexlab#1{#1}\fi
\expandafter\ifx\csname bibnamefont\endcsname\relax
  \def\bibnamefont#1{#1}\fi
\expandafter\ifx\csname bibfnamefont\endcsname\relax
  \def\bibfnamefont#1{#1}\fi
\expandafter\ifx\csname citenamefont\endcsname\relax
  \def\citenamefont#1{#1}\fi
\expandafter\ifx\csname url\endcsname\relax
  \def\url#1{\texttt{#1}}\fi
\expandafter\ifx\csname urlprefix\endcsname\relax\def\urlprefix{URL }\fi
\providecommand{\bibinfo}[2]{#2}
\providecommand{\eprint}[2][]{\url{#2}}

\bibitem[{\citenamefont{Alava et~al.}(2006)\citenamefont{Alava, Nukala, and
  Zapperi}}]{Alava2}
\bibinfo{author}{\bibfnamefont{M.~J.} \bibnamefont{Alava}},
  \bibinfo{author}{\bibfnamefont{P.~K.} \bibnamefont{Nukala}},
  \bibnamefont{and} \bibinfo{author}{\bibfnamefont{S.}~\bibnamefont{Zapperi}},
  \bibinfo{journal}{Adv. Phys.} \textbf{\bibinfo{volume}{55}},
  \bibinfo{pages}{349} (\bibinfo{year}{2006}).

\bibitem[{\citenamefont{M{\aa}l{\o}y and Schmittbuhl}(2001)}]{Maloy2}
\bibinfo{author}{\bibfnamefont{K.~J.} \bibnamefont{M{\aa}l{\o}y}}
  \bibnamefont{and}
  \bibinfo{author}{\bibfnamefont{J.}~\bibnamefont{Schmittbuhl}},
  \bibinfo{journal}{Phys. Rev. Lett.} \textbf{\bibinfo{volume}{87}},
  \bibinfo{pages}{105502} (\bibinfo{year}{2001}).

\bibitem[{\citenamefont{M{\aa}l{\o}y et~al.}(2006)\citenamefont{M{\aa}l{\o}y,
  Santucci, Schmittbuhl, and Toussaint}}]{Maloy3}
\bibinfo{author}{\bibfnamefont{K.~J.} \bibnamefont{M{\aa}l{\o}y}},
  \bibinfo{author}{\bibfnamefont{S.}~\bibnamefont{Santucci}},
  \bibinfo{author}{\bibfnamefont{J.}~\bibnamefont{Schmittbuhl}},
  \bibnamefont{and}
  \bibinfo{author}{\bibfnamefont{R.}~\bibnamefont{Toussaint}},
  \bibinfo{journal}{Phys. Rev. Lett.} \textbf{\bibinfo{volume}{96}},
  \bibinfo{pages}{045501} (\bibinfo{year}{2006}).

\bibitem[{\citenamefont{Koivisto et~al.}(2007)\citenamefont{Koivisto, Rosti,
  and Alava}}]{Koivisto}
\bibinfo{author}{\bibfnamefont{J.}~\bibnamefont{Koivisto}},
  \bibinfo{author}{\bibfnamefont{J.}~\bibnamefont{Rosti}}, \bibnamefont{and}
  \bibinfo{author}{\bibfnamefont{M.~J.} \bibnamefont{Alava}},
  \bibinfo{journal}{Phys. Rev. Lett.} \textbf{\bibinfo{volume}{99}},
  \bibinfo{pages}{145504} (\bibinfo{year}{2007}).

\bibitem[{\citenamefont{Davidsen et~al.}(2007)\citenamefont{Davidsen,
  Stanchits, and Dresen}}]{Davidsen}
\bibinfo{author}{\bibfnamefont{J.}~\bibnamefont{Davidsen}},
  \bibinfo{author}{\bibfnamefont{S.}~\bibnamefont{Stanchits}},
  \bibnamefont{and} \bibinfo{author}{\bibfnamefont{G.}~\bibnamefont{Dresen}},
  \bibinfo{journal}{Phys. Rev. Lett.} \textbf{\bibinfo{volume}{98}},
  \bibinfo{pages}{125502} (\bibinfo{year}{2007}).

\bibitem[{\citenamefont{Bouchaud et~al.}(1990)\citenamefont{Bouchaud, Lapasset,
  and Plan\`es}}]{Bouchaud9}
\bibinfo{author}{\bibfnamefont{E.}~\bibnamefont{Bouchaud}},
  \bibinfo{author}{\bibfnamefont{G.}~\bibnamefont{Lapasset}}, \bibnamefont{and}
  \bibinfo{author}{\bibfnamefont{J.}~\bibnamefont{Plan\`es}},
  \bibinfo{journal}{Europhys. Lett.} \textbf{\bibinfo{volume}{13}},
  \bibinfo{pages}{73} (\bibinfo{year}{1990}).

\bibitem[{\citenamefont{M{\aa}l{\o}y et~al.}(1992)\citenamefont{M{\aa}l{\o}y,
  Hansen, Hinrichsen, and Roux}}]{Maloy}
\bibinfo{author}{\bibfnamefont{K.~J.} \bibnamefont{M{\aa}l{\o}y}},
  \bibinfo{author}{\bibfnamefont{A.}~\bibnamefont{Hansen}},
  \bibinfo{author}{\bibfnamefont{E.~L.} \bibnamefont{Hinrichsen}},
  \bibnamefont{and} \bibinfo{author}{\bibfnamefont{S.}~\bibnamefont{Roux}},
  \bibinfo{journal}{Phys. Rev. Lett.} \textbf{\bibinfo{volume}{68}},
  \bibinfo{pages}{213} (\bibinfo{year}{1992}).

\bibitem[{\citenamefont{Ponson et~al.}(2006)\citenamefont{Ponson, Bonamy, and
  Bouchaud}}]{Ponson5}
\bibinfo{author}{\bibfnamefont{L.}~\bibnamefont{Ponson}},
  \bibinfo{author}{\bibfnamefont{D.}~\bibnamefont{Bonamy}}, \bibnamefont{and}
  \bibinfo{author}{\bibfnamefont{E.}~\bibnamefont{Bouchaud}},
  \bibinfo{journal}{Phys. Rev. Lett.} \textbf{\bibinfo{volume}{96}},
  \bibinfo{pages}{035506} (\bibinfo{year}{2006}).

\bibitem[{\citenamefont{Bonamy et~al.}(2006)\citenamefont{Bonamy, Ponson,
  Prades, bouchaud, and Guillot}}]{Bonamy2}
\bibinfo{author}{\bibfnamefont{D.}~\bibnamefont{Bonamy}},
  \bibinfo{author}{\bibfnamefont{L.}~\bibnamefont{Ponson}},
  \bibinfo{author}{\bibfnamefont{S.}~\bibnamefont{Prades}},
  \bibinfo{author}{\bibfnamefont{E.}~\bibnamefont{bouchaud}}, \bibnamefont{and}
  \bibinfo{author}{\bibfnamefont{C.}~\bibnamefont{Guillot}},
  \bibinfo{journal}{Phys. Rev. Lett.} \textbf{\bibinfo{volume}{97}},
  \bibinfo{pages}{135504} (\bibinfo{year}{2006}).

\bibitem[{\citenamefont{Bonamy et~al.}(2008)\citenamefont{Bonamy, Santucci, and
  Ponson}}]{Bonamy5}
\bibinfo{author}{\bibfnamefont{D.}~\bibnamefont{Bonamy}},
  \bibinfo{author}{\bibfnamefont{S.}~\bibnamefont{Santucci}}, \bibnamefont{and}
  \bibinfo{author}{\bibfnamefont{L.}~\bibnamefont{Ponson}},
  \bibinfo{journal}{cond-mat/0803.0190}  (\bibinfo{year}{2008}).

\bibitem[{\citenamefont{Lawn}(1993)}]{Lawn}
\bibinfo{author}{\bibfnamefont{B.}~\bibnamefont{Lawn}},
  \emph{\bibinfo{title}{Fracture of brittle solids}}
  (\bibinfo{publisher}{Cambridge University Press}, \bibinfo{year}{1993}).

\bibitem[{\citenamefont{Schmittbuhl
  et~al.}(1995{\natexlab{a}})\citenamefont{Schmittbuhl, Roux, Vilotte, and
  M{\aa}l{\o}y}}]{Schmittbuhl4}
\bibinfo{author}{\bibfnamefont{J.}~\bibnamefont{Schmittbuhl}},
  \bibinfo{author}{\bibfnamefont{S.}~\bibnamefont{Roux}},
  \bibinfo{author}{\bibfnamefont{J.~P.} \bibnamefont{Vilotte}},
  \bibnamefont{and} \bibinfo{author}{\bibfnamefont{K.~J.}
  \bibnamefont{M{\aa}l{\o}y}}, \bibinfo{journal}{Phys. Rev. Lett.}
  \textbf{\bibinfo{volume}{74}}, \bibinfo{pages}{1787}
  (\bibinfo{year}{1995}{\natexlab{a}}).

\bibitem[{\citenamefont{Ramanathan et~al.}(1997)\citenamefont{Ramanathan,
  Ertas, and Fisher}}]{Ramanathan}
\bibinfo{author}{\bibfnamefont{S.}~\bibnamefont{Ramanathan}},
  \bibinfo{author}{\bibfnamefont{D.}~\bibnamefont{Ertas}}, \bibnamefont{and}
  \bibinfo{author}{\bibfnamefont{D.~S.} \bibnamefont{Fisher}},
  \bibinfo{journal}{Phys. Rev. Lett.} \textbf{\bibinfo{volume}{79}},
  \bibinfo{pages}{873} (\bibinfo{year}{1997}).

\bibitem[{\citenamefont{Atkinson}(1984)}]{Atkinson}
\bibinfo{author}{\bibfnamefont{B.~K.} \bibnamefont{Atkinson}},
  \bibinfo{journal}{J. Geophys. Res.} \textbf{\bibinfo{volume}{89}},
  \bibinfo{pages}{4077} (\bibinfo{year}{1984}).

\bibitem[{\citenamefont{Carneiro and Barcellos}(1953)}]{Carneiro2}
\bibinfo{author}{\bibfnamefont{F.~L. L.~B.} \bibnamefont{Carneiro}}
  \bibnamefont{and}
  \bibinfo{author}{\bibfnamefont{A.}~\bibnamefont{Barcellos}},
  \bibinfo{journal}{Bulletin RILEM} \textbf{\bibinfo{volume}{1}},
  \bibinfo{pages}{97} (\bibinfo{year}{1953}).

\bibitem[{\citenamefont{Barenblatt}(1962)}]{Barenblatt}
\bibinfo{author}{\bibfnamefont{G.~I.} \bibnamefont{Barenblatt}},
  \bibinfo{journal}{Adv. Appl. Mech.} \textbf{\bibinfo{volume}{7}},
  \bibinfo{pages}{55} (\bibinfo{year}{1962}).

\bibitem[{\citenamefont{Evans}(1972)}]{Evans}
\bibinfo{author}{\bibfnamefont{A.~G.} \bibnamefont{Evans}},
  \bibinfo{journal}{J. Mater. Sci.} \textbf{\bibinfo{volume}{15}},
  \bibinfo{pages}{1137} (\bibinfo{year}{1972}).

\bibitem[{\citenamefont{Nara and Kaneko}(2005)}]{Nara}
\bibinfo{author}{\bibfnamefont{Y.}~\bibnamefont{Nara}} \bibnamefont{and}
  \bibinfo{author}{\bibfnamefont{K.}~\bibnamefont{Kaneko}},
  \bibinfo{journal}{Int. J. Rock. Mech. Min. Sci.} pp.
  \bibinfo{pages}{521--530} (\bibinfo{year}{2005}).

\bibitem[{\citenamefont{Note1}(2005)}]{Note1}
The measurement of the test machine work is made from the displacement $\delta^\indice{F}(t) = \beta^\indice{FE}(c)\,\delta(t) $ between the points of application of forces where $\beta^\indice{FE}(c) \simeq 0.8$ is provided by FE simulations.

\bibitem[{\citenamefont{Morel et~al.}(2005)\citenamefont{Morel, Dourado,
  Valentin, and Morais}}]{Morel7}
\bibinfo{author}{\bibfnamefont{S.}~\bibnamefont{Morel}},
  \bibinfo{author}{\bibfnamefont{N.}~\bibnamefont{Dourado}},
  \bibinfo{author}{\bibfnamefont{G.}~\bibnamefont{Valentin}}, \bibnamefont{and}
  \bibinfo{author}{\bibfnamefont{J.}~\bibnamefont{Morais}},
  \bibinfo{journal}{Int. J. Frac.} \textbf{\bibinfo{volume}{131}},
  \bibinfo{pages}{385} (\bibinfo{year}{2005}).

\bibitem[{\citenamefont{Charles}(1959)}]{Charles}
\bibinfo{author}{\bibfnamefont{R.~J.} \bibnamefont{Charles}},
  \bibinfo{journal}{J. Appl. Phys.} \textbf{\bibinfo{volume}{29}},
  \bibinfo{pages}{1554} (\bibinfo{year}{1959}).

\bibitem[{\citenamefont{Wiederhorn et~al.}(1974)\citenamefont{Wiederhorn,
  Johnson, Dinessand, and Heuer}}]{Wiederhorn2}
\bibinfo{author}{\bibfnamefont{S.~M.} \bibnamefont{Wiederhorn}},
  \bibinfo{author}{\bibfnamefont{H.}~\bibnamefont{Johnson}},
  \bibinfo{author}{\bibfnamefont{A.~M.} \bibnamefont{Dinessand}},
  \bibnamefont{and} \bibinfo{author}{\bibfnamefont{A.~H.} \bibnamefont{Heuer}},
  \bibinfo{journal}{J. Am. Ceram. Soc.} \textbf{\bibinfo{volume}{57}}
  (\bibinfo{year}{1974}).

\bibitem[{\citenamefont{Santucci et~al.}(2004)\citenamefont{Santucci, Vanel,
  and Ciliberto}}]{Santucci3}
\bibinfo{author}{\bibfnamefont{S.}~\bibnamefont{Santucci}},
  \bibinfo{author}{\bibfnamefont{L.}~\bibnamefont{Vanel}}, \bibnamefont{and}
  \bibinfo{author}{\bibfnamefont{S.}~\bibnamefont{Ciliberto}},
  \bibinfo{journal}{Phys. Rev. Lett.} \textbf{\bibinfo{volume}{93}},
  \bibinfo{pages}{095505} (\bibinfo{year}{2004}).

\bibitem[{\citenamefont{Wiederhorn}(1967)}]{Wiederhorn}
\bibinfo{author}{\bibfnamefont{S.~M.} \bibnamefont{Wiederhorn}},
  \bibinfo{journal}{J. Am. Ceram. Soc.} \textbf{\bibinfo{volume}{50}},
  \bibinfo{pages}{407} (\bibinfo{year}{1967}).

\bibitem[{\citenamefont{Ponson et~al.}(2007)\citenamefont{Ponson, Bonamy,
  Bouchaud, Cordeiro, Toledo, and Fairbairn}}]{Ponson13}
\bibinfo{author}{\bibfnamefont{L.}~\bibnamefont{Ponson}},
  \bibinfo{author}{\bibfnamefont{D.}~\bibnamefont{Bonamy}},
  \bibinfo{author}{\bibfnamefont{E.}~\bibnamefont{Bouchaud}},
  \bibinfo{author}{\bibfnamefont{G.}~\bibnamefont{Cordeiro}},
  \bibinfo{author}{\bibfnamefont{R.}~\bibnamefont{Toledo}}, \bibnamefont{and}
  \bibinfo{author}{\bibfnamefont{E.}~\bibnamefont{Fairbairn}}, in
  \emph{\bibinfo{booktitle}{Proceeding FRAMCOS-6}}, edited by
  \bibinfo{editor}{\bibfnamefont{A.}~\bibnamefont{Carpinteri}}
  \bibnamefont{and} \bibinfo{editor}{\bibnamefont{{\em et al.}}}
  (\bibinfo{year}{2007}), pp. \bibinfo{pages}{63--67}.

\bibitem[{\citenamefont{Mochvan and Willis}(2000)}]{Mochvan}
\bibinfo{author}{\bibfnamefont{A.~B.} \bibnamefont{Mochvan}} \bibnamefont{and}
  \bibinfo{author}{\bibfnamefont{J.~R.} \bibnamefont{Willis}},
  \bibinfo{journal}{J. Eng. Math.} \textbf{\bibinfo{volume}{37}},
  \bibinfo{pages}{143} (\bibinfo{year}{2000}).

\bibitem[{\citenamefont{Ball and Larralde}(1995)}]{Ball}
\bibinfo{author}{\bibfnamefont{R.~C.} \bibnamefont{Ball}} \bibnamefont{and}
  \bibinfo{author}{\bibfnamefont{H.}~\bibnamefont{Larralde}},
  \bibinfo{journal}{Int. J. Frac.} \textbf{\bibinfo{volume}{71}},
  \bibinfo{pages}{365} (\bibinfo{year}{1995}).

\bibitem[{\citenamefont{Rice}(1985)}]{Rice}
\bibinfo{author}{\bibfnamefont{J.~R.} \bibnamefont{Rice}},
  \bibinfo{journal}{J. Appl. Mech.} \textbf{\bibinfo{volume}{52}},
  \bibinfo{pages}{571} (\bibinfo{year}{1985}).

\bibitem[{\citenamefont{Barab\'asi and Stanley}(1995)}]{Barabasi}
\bibinfo{author}{\bibfnamefont{A.~L.} \bibnamefont{Barab\'asi}}
  \bibnamefont{and} \bibinfo{author}{\bibfnamefont{H.~E.}
  \bibnamefont{Stanley}}, \emph{\bibinfo{title}{Fractal concepts in surface
  growth}} (\bibinfo{publisher}{Cambridge University Press},
  \bibinfo{year}{1995}).

\bibitem[{\citenamefont{Ertas and Kardar}(1994)}]{Ertas2}
\bibinfo{author}{\bibfnamefont{D.}~\bibnamefont{Ertas}} \bibnamefont{and}
  \bibinfo{author}{\bibfnamefont{M.}~\bibnamefont{Kardar}},
  \bibinfo{journal}{Phys. Rev. E} \textbf{\bibinfo{volume}{49}},
  \bibinfo{pages}{R2532} (\bibinfo{year}{1994}).

\bibitem[{\citenamefont{Duemmer and Krauth}(2007)}]{Duemmer3}
\bibinfo{author}{\bibfnamefont{O.}~\bibnamefont{Duemmer}} \bibnamefont{and}
  \bibinfo{author}{\bibfnamefont{W.}~\bibnamefont{Krauth}},
  \bibinfo{journal}{J. Stat. Mech.} \textbf{\bibinfo{volume}{1}},
  \bibinfo{pages}{01019} (\bibinfo{year}{2007}).

\bibitem[{\citenamefont{Feigelman et~al.}(1989)\citenamefont{Feigelman,
  Geshkenbein, Larkin, and Vinokur}}]{Feigelman}
\bibinfo{author}{\bibfnamefont{M.~V.} \bibnamefont{Feigelman}},
  \bibinfo{author}{\bibfnamefont{V.~B.} \bibnamefont{Geshkenbein}},
  \bibinfo{author}{\bibfnamefont{A.~I.} \bibnamefont{Larkin}},
  \bibnamefont{and} \bibinfo{author}{\bibfnamefont{V.~M.}
  \bibnamefont{Vinokur}}, \bibinfo{journal}{Phys. Rev. Lett.}
  \textbf{\bibinfo{volume}{63}}, \bibinfo{pages}{2303} (\bibinfo{year}{1989}).

\bibitem[{\citenamefont{Kolton et~al.}(2005)\citenamefont{Kolton, Rosso, and
  Giamarchi}}]{Kolton}
\bibinfo{author}{\bibfnamefont{A.~B.} \bibnamefont{Kolton}},
  \bibinfo{author}{\bibfnamefont{A.}~\bibnamefont{Rosso}}, \bibnamefont{and}
  \bibinfo{author}{\bibfnamefont{T.}~\bibnamefont{Giamarchi}},
  \bibinfo{journal}{Phys. Rev. Lett.} \textbf{\bibinfo{volume}{94}},
  \bibinfo{pages}{047002} (\bibinfo{year}{2005}).

\end{thebibliography}
\end{document}